\documentclass[aps,prb,twocolumn,reprint,showpacs]{revtex4-1}
\usepackage{amsmath}
\usepackage{amssymb}
\usepackage{graphicx}
\usepackage{hyperref}
\usepackage{url}
\usepackage{color,xcolor}
\usepackage{epstopdf}

\begin{document}
\title{Protective layer enhanced stability and superconductivity of tailored antimonene bilayer}
\author{Jun-Jie Zhang}
\author{Yang Zhang}
\author{Shuai Dong}
\email{sdong@seu.edu.cn}
\affiliation{School of Physics, Southeast University, Nanjing 211189, China}
\date{\today}

\begin{abstract}
For two-dimensional superconductors, the high stability in ambient conditions is critical for experiments and applications. Few-layer antimonene can be non-degradative over a couple of months, which is superior to the akin black phosphorus. Based on the anisotropic Migdal-Eliashberg theory and maximally-localised Wannier functions, this work predicts that electron-doping and Ca-intercalation can transform $\beta$-Sb bilayer from a semimetal to a superconductor. However, the stability of antimonene bilayer in air trends to be decreased due to the electron doping. To overcome this drawback, two kinds of protective layers (graphene and $h$-BN) are proposed to enhance the stability. Interestingly, the superconducting transition temperature will also be enhanced to $9.6$ K, making it a promising candidate as nanoscale superconductor.
\end{abstract}
\maketitle

\section{Introduction}
Recently, few-layer black phosphorus (BP) as an emerging member of two-dimensional (2D) materials has drawn lots of research attentions due to its unique properties, such as high carrier mobility (up to $1000$ cm$^2$V$^{-1}$s$^{-1}$ at room temperature) \cite{qiao2014high} and anisotropic in-plane properties \cite{tran2014layer}. However, the stability of few-layer BP remains a big challenge in experiments, on account of decomposition in air within several hours \cite{castellanos2014isolation,island2015environmental}. Antimony, as another member of nitrogen group, has several 2D phases (referred to as antimonene), such as $\alpha$-, $\beta$-, $\gamma$-, and $\delta$-Sb \cite{zhang2016semiconducting}. Among these allotropes, the $\beta$-Sb owns the lowest energy \cite{zhang2016semiconducting}, which has been successfully prepared in recent experiments \cite{ares2016mechanical,ji2016two}. Comparing with black phosphorus, few-layer antimonene is more stable in ambient conditions, which can be non-degraded over a couple of months \cite{ares2016mechanical,ji2016two}. Interestingly, $\beta$-Sb transforms from a semimetal to a semiconductor with decreasing thickness to few layers, and the monolayer shows an indirect band gap of $\sim2$ eV \cite{zhang2016semiconducting,kamal2015arsenene}. It is also predicted that antimonene owns high carrier mobility \cite{zhang2016semiconducting}, wide range optical absorption \cite{ares2016mechanical}, and high refractive index \cite{singh2016antimonene}, which may lead to potential applications in optoelectronics. Thus, lots of experimental and theoretical works have been done on antimonene recently.

Superconductivity in 2D materials has also attracted enormous interests \cite{ge2015superconductivity}. Recently, Penev \textit{et. al.} predicted that 2D boron monolayer would exhibit intrinsic phonon-mediated superconductivity \cite{penev2016can}. Our recent study also predicted Mo$_{2}$C monolayer to be a quasi-2D superconductor, whose superconducting transition temperature ($T_{\rm C}$) could be adjusted by functional groups \cite{zhang2017superconductivity}. Moreover, superconductivity could also be induced and enhanced in other semiconducting or semimetal 2D materials via various methods, e.g. by improving carrier density and electron-phonon coupling (EPC) \cite{zhang2016strain,zhang2016blue,chi2015ultrahigh}. Even though, the high stability in ambient conditions of 2D superconductors is critical for experiments and applications, which has been rarely touched in these theoretical studies. In this sense, tailored antimonene may be a good candidate for 2D superconductor. Thus, it is highly desired to carefully investigate the stability and superconductivity of tailored antimonene.

In this work, using density functional theory (DFT), anisotropic Migdal-Eliashberg theory, and maximally-localised Wannier functions (MLWFs), the stability and superconductivity of electron-doped and Ca-intercalated $\beta$-Sb bilayer have been investigated. Our calculation finds that it will transform $\beta$-Sb bilayer from a semimetal to superconductor by increasing the carrier density. However, when the doping density is $>0.8$ electrons/cell, the strong EPC would lead to dynamic instability of the $\beta$-Sb bilayer structure. To pursuit stable and superconducting $\beta$-Sb bilayer, two kinds of protective layers, graphene and $h$-BN, have been considered, which can improve the superconducting $T_{\rm C}$ to $9.6$ K.

\section{Model \& methods}
The electronic structure, molecular dynamics (MD), and climbing-image nudged elastic band (CI-NEB) calculations have been performed using the Vienna {\it ab initio} simulation package (VASP) with projector-augmented wave (PAW) potentials \cite{kresse1996efficient,kresse1999ultrasoft}. The generalized gradient approximation of Perdew-Burke-Ernzerhof (GGA-PBE) formulation are used with a cutoff energy $500$ eV. The phonon dispersion and electron-phonon coupling calculations are carried out using the Quantum-ESPRESSO distribution, which are calculated within the framework of density function perturbation theory (DFPT) \cite{giannozzi2009quantum}. The full relativistic norm-conserving potential are generated by $ld1.x$, which have been well tested for further calculations. The cutoff energy for the expanding wave function is $65$ Ry. Structure optimization and electronic structure are repeated by using Quantum-ESPRESSO and the results are consistent with those obtained using VASP. In order to deeply understand the electronic and superconducting properies, the recently developed Wannier interpolation technique is also used in our calculation \cite{giustino2007electron,noffsinger2010epw}. More computational details are given in the Supplementary Materials \cite{Supp}.

\section{Results \& discussion}
\subsection{Pristine antimonene bilayer}
Among all known allotropes of antimonene, $\beta$-Sb is the ground state of monolayers \cite{zhang2016semiconducting}. Different from planar graphene/$h$-BN, the monolayer $\beta$-Sb owns the buckled structure due to $sp^3$ hybridization. The calculated electronic structure and phonon dispersion of $\beta$-Sb monolayer are shown in Fig.~S1 \cite{Supp}, in agreement with previous studies \cite{zhang2015atomically,akturk2015single,wang2015atomically}.

\begin{figure}
\centering
\includegraphics[width=0.4\textwidth]{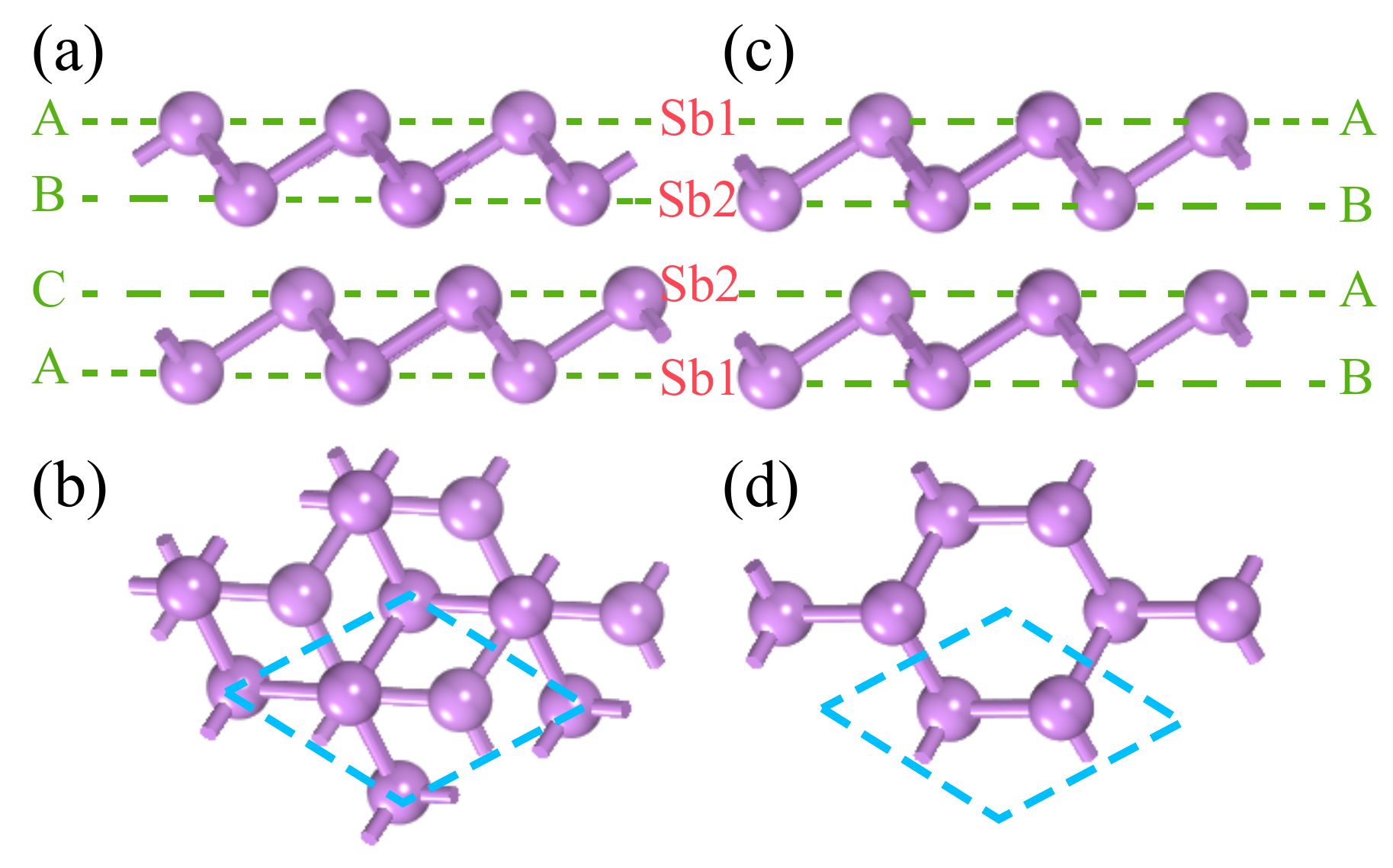}
\caption{Side and top views of atomic structures of $\beta$-Sb bilayer with different stacking configurations. (a) A-B CG; (b) A-A CG.}
\label{Fig1}
\end{figure}

The $\beta$-Sb bilayer adopts the A-B type configuration (CG) [Fig.~\ref{Fig1}(a)], i.e. the stacking sequence of atomic planes is $AB~CA$ \cite{zhang2016semiconducting}, while the A-A CG with a mirror symmetry between two layers [Fig.~\ref{Fig1}(b)] owns a little higher energy (about $12.5$ meV per atom) than the A-B CG. These two CGs own the same point group, i.e. $D_{3d}$ with $12$ symmetry operations, and both are dynamically stable according to their phonon spectrums [Fig.~S2(a-b)]. The optimized lattice constants are $4.142$ {\AA} and $4.075$ {\AA} for A-B and A-A CG, respectively.

\begin{figure}
\centering
\includegraphics[width=0.45\textwidth]{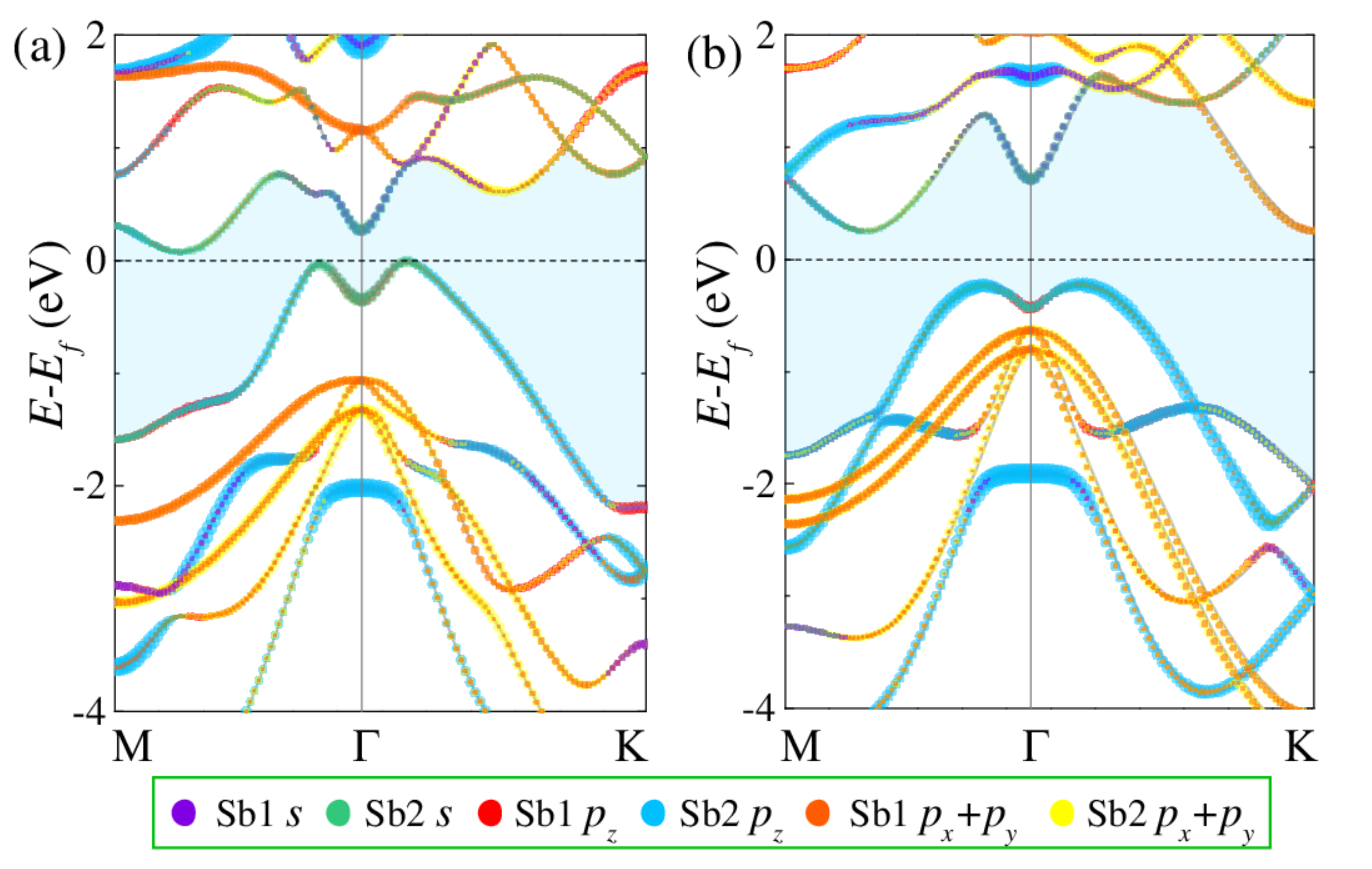}
\caption{Projected electronic band structures of $\beta$-Sb bilayer with different stacking. (a) A-B CG; (b) A-A CG.}
\label{Fig2}
\end{figure}

From their corresponding Raman intensity [Fig.~S2(c)], the interlayer breathing mode ($A_{\rm 1g}^1$) and in-plane shear mode ($E_{\rm g}^1$) of $\beta$-Sb bilayer locate in low frequency range (less than $50$ cm$^{-1}$), due to the weak interlayer interaction. Since the calculated interlayer distance ($d$) is shorter in the A-B CG ($2.39$ {\AA} for A-B and $3.07$ {\AA} for A-A), the stronger interlayer interaction leads to harder breathing mode and in-plane shear mode. The results imply that considerable proportion of chemical interaction between layers in the A-B CG, beyond the pure vdW interaction \cite{akturk2015single}. Such chemical interaction leads to significant changes in the electronic structure of $\beta$-Sb bilayer, as shown in Fig.~\ref{Fig2}. Specifically, the band structure of A-B CG [Fig.~\ref{Fig2}(a)] shows the semimetallic characters (band gap $\sim0$ eV), while the A-A CG remains an indirect semiconductor with a gap of $\sim0.4$ eV. The valence band maximum (VBM) and conduction band minimum (CBM) in the A-A and A-B CG's are mainly contributed by Sb2's $s$- and $p_z$-orbitals. Furthermore, using the MLWFs, the nearest-neighbor interlayer hopping of $p_z$ orbitals reaches $0.41$ eV, a considerable magnitude. To further verify the crucial role of interlayer interactions in the electronic structure, the interlayer distance is artificially lifted to $3.0$ {\AA} and $3.5$ {\AA}, which gives rise to finite band gaps (Fig.~S3).

\subsection{Superconductivity of tailored antimonene bilayer}
\begin{figure}
\centering
\includegraphics[width=0.43\textwidth]{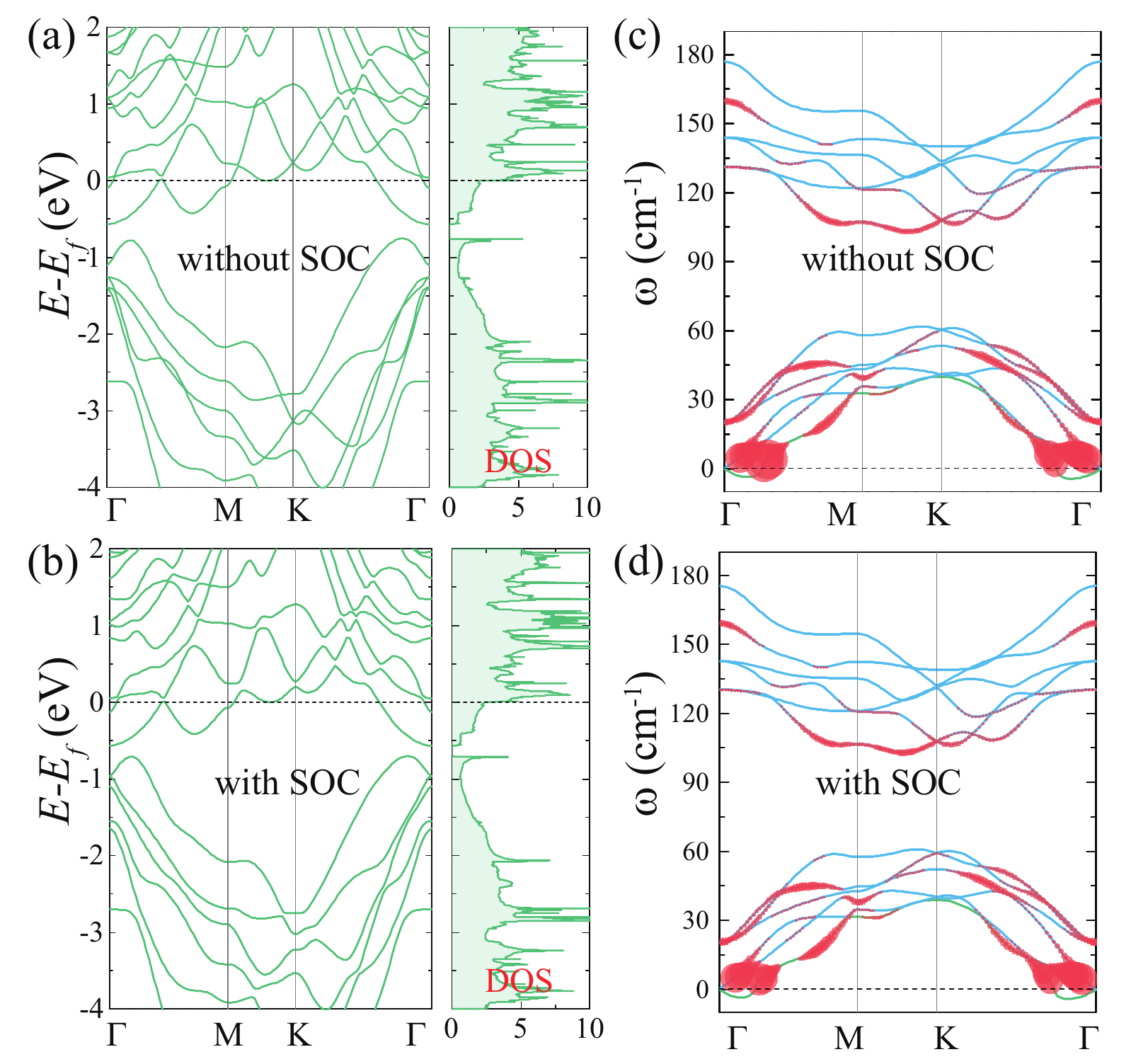}
\caption{$\beta$-Sb bilayer with 1.0 electron/cell doping concentration. (a) Electronic band structure. (b) Phonon dispersion. The red circles indicate the strength of electron-phonon coupling.}
\label{Fig3new1}
\end{figure}

Doping is a powerful method to improve carrier density in materials. To verify the maximum doping concentrations in $\beta$-Sb bilayer, pure electron doping will be tested first. Five doping concentrations ($0.2$, $0.4$, $0.6$, $0.8$, and $1.0$ electrons/cell) are considered here, and the corresponding maximal carrier density is about $6.6\times10^{14}$ cm$^{-1}$. After doping, the in-plane lattice constants and atomic positions are further optimized, and the A-B CG still owns the lowest energy. As expected, electron doping makes $\beta$-Sb bilayer transform from a semimetal to a metal [Fig.~S4 and Fig.~\ref{Fig3new1}(a-b)]. It should be noted that the spin-orbit coupling (SOC) has very limited influence on the electron structure of electron-doped A-B CG [Fig.~S4 and Fig.~\ref{Fig3new1}(a-b)], although Sb is a heavy element.

%According to the band structure (Fig.~S4), the doping makes $\beta$-Sb bilayer transform from a semimetal to metal. %The conduction band is pushed down in $\Gamma$-M point, and the corresponding Fermi surface is enlarged by electron doping. When the doping concentration is increased to $0.4$ electrons/cell, the second lowest conduction band starts crossing the Fermi level near $\Gamma$ point, and a new cyclic-shape Fermi surface starts to show up.
%It should be noted that with the $D_{3d}$ group of A-B CG, the bands are obviously doubly degenerated under time- and inversion-reversal symmetries. Thus the SOC can not lead to spin splitting, i.e., the SOC has a limited influence on the electron structure of electron-doped A-B CG (Fig.~S4), although Sb is a heavy element.%

The calculated phonon dispersions and EPC constant $\lambda_{\bf{q}\nu}$ for electron-doped antimonene bilayer are shown in Fig.~S5. In addition, the detailed definition and calculation descriptions of EPC are given in the Supplementary Materials \cite{Supp}. %In Fig.~S5, one can notice that all Raman modes are slightly softened upon doping, as summarized in Fig.~S6. Such softening behavior generally exists in some phonon-mediated superconductors, e.g. Mg$B_{2}$.\cite{kortus2001superconductivity} Most importantly, the ${\bf q}$ of strong $\lambda_{\bf{q}\nu}$ matches the Fermi vectors around $\Gamma$ and $M$ points, which is a typical signature of Fermi contour nesting. Specially, among the optical modes, the breathing mode $A_{1g}^1$ and intralayer mode $E_{g}^2$ have the strongest coupling to electron.
When the doping concentration is up to $1.0$ electrons/cell, there is a quadratic phonon of the $ZA$ mode (out-of-plane acoustical mode) becoming imaginary round the $\Gamma$ point due to the strong EPC [Fig.~\ref{Fig3new1}(c-d)]. According to the density of states (DOS) [Fig.~\ref{Fig3new1}(a-b)], the Fermi level at $1.0$ electrons/cell doping is moved to near the Van Hove singular point, resulting in the structural instability. Therefore, the maximum doping concentration is around $7.0\times10^{14}$ cm$^{-1}$ for free-standing $\beta$-Sb bilayer. The biaxial tension can remove this negative frequency around the long-wavelength limit, where $ZA$ mode becomes linear dispersion ($\omega\sim|{\bf q}|$) in the vicinity of $\Gamma$ point \cite{penev2016can,zhou2015high}.

\begin{figure}
\centering
\includegraphics[width=0.48\textwidth]{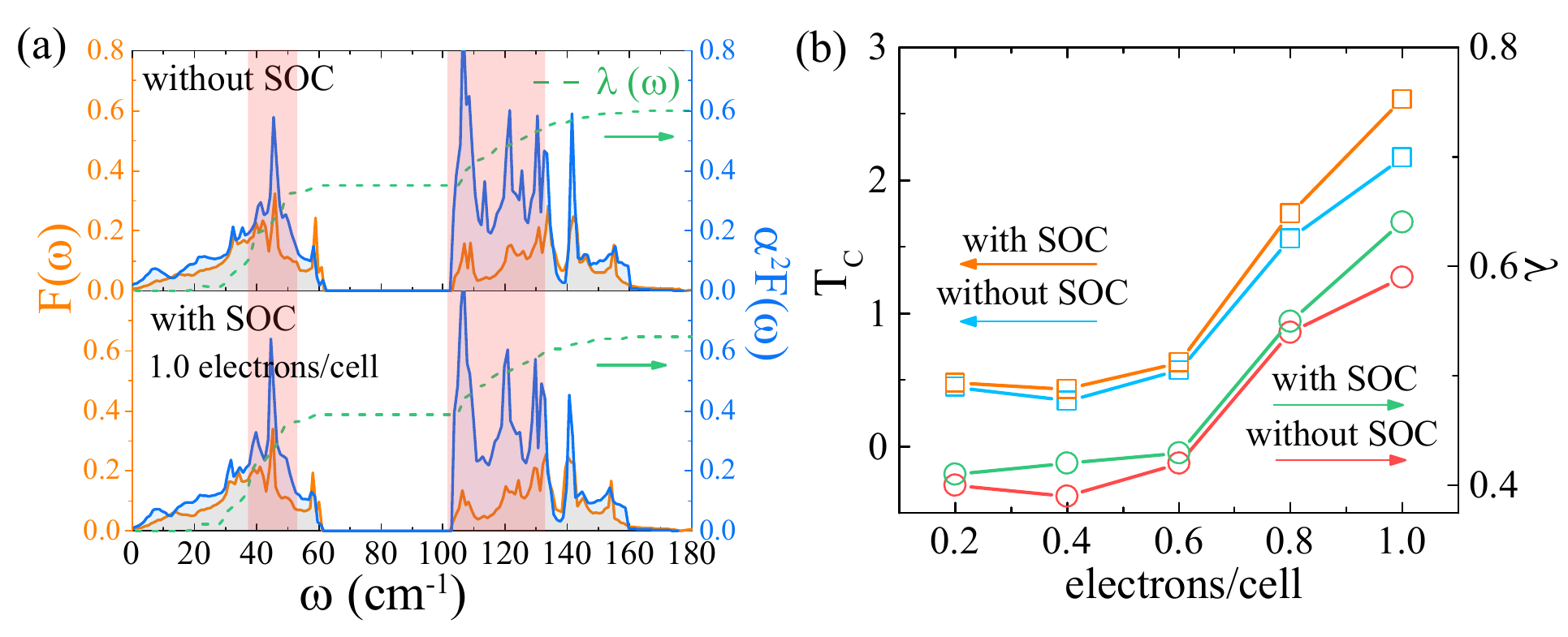}
\caption{(a) Phonon density of states (PDOS) $F(\omega)$, Eliashberg-spectral function $\alpha^{2}F(\omega)$, and electron-phonon coupling $\lambda(\omega)$ with 1.0 electron/cell doping concentration. (b) Electron-phonon coupling constant $\lambda$ and superconducting $T_{\rm C}$ under different doping concentration.}
\label{Fig3new}
\end{figure}

To study the superconductivity of electron-doped $\beta$-Sb bilayer, the EPC constant $\lambda$ is also calculated by summing over the first Brillouin zone (BZ), or integrating the $\alpha^{2}F(\omega)$ in the $\bf{q}$ space \cite{allen1975transition,allen1972neutron}:
\begin{equation}
\lambda=\sum_{\bf{q}\nu}\lambda_{\bf{q}\nu}=2\int_{0}^{\infty}\frac{\alpha^2F(\omega)}{\omega}d\omega.
\label{lambda}
\end{equation}
The obtained $\alpha^{2}F(\omega)$, phonon density of states (PDOS, $F(\omega)$) and $\lambda(\omega)$ are plotted in Fig.~\ref{Fig3new}(a) and Fig.~S6. Comparing with $F(\omega)$'s, all $\alpha^{2}F(\omega)$'s have similar shape, implying that all vibration modes contribute to the EPC and corresponding frequency region is very remarkable (red area of Fig.~\ref{Fig3new}(a) and Fig.~S6). Due to the factor $1/\omega$ in the definition of $\lambda$ (see Eq.~\ref{lambda}), the contributions from high-frequency is negligible, which are apparently shown in Fig.~\ref{Fig3new}(a) and Fig.~S6. %For instance, when doping concentration is $0.2$ electrons/cell, the calculated $\lambda(\omega=73$ cm$^{-1})$ is $\approx0.26$ which is beyond $65\%$ of the total EPC ($\lambda(\omega=\infty)=0.40$), indicating that the phonon modes in the frequency region below $73$ cm$^{-1}$ have the dominant contribution.%
Exactly, the frequency regions of $A_{\rm 1g}^1$ and $E_{\rm g}^2$ correspond to area-A and area-B, respectively. The obvious peaks of $\alpha^{2}F(\omega)$ exist in these two areas, indicating that these two vibrational modes have great effects on the superconductivity of electron-doped antimonene bilayer. The $T_{\rm C}$ can be estimated using the Allen-Dynes modified McMillan equation \cite{allen1975transition}:
\begin{equation}
T_{\rm C}=\frac{\omega_{ln}}{1.2}\exp[-\frac{1.04(1+\lambda)}{\lambda-\mu^*(1+0.62\lambda)}],
\label{McMillan}
\end{equation}
where $\mu^*$ is the Coulomb repulsion parameter and $\omega_{ln}$ is the logarithmically averaged frequency. When taking a typical value $\mu^{*}=0.1$, the estimated $T_{\rm C}$ and corresponding superconductive parameter are shown in Fig.~\ref{Fig3new}(b). When the doping concentration is less than $0.6$ electrons/cell, all $T_{\rm C}$'s are under $1$ K, and the effect of SOC is not obvious. The EPC is rapidly enhanced when more electrons are added to $\beta$-Sb bilayer and the SOC may also promote EPC. The highest $T_{\rm C}$ can reach $2.6$ K.

\begin{figure}
\centering
\includegraphics[width=0.45\textwidth]{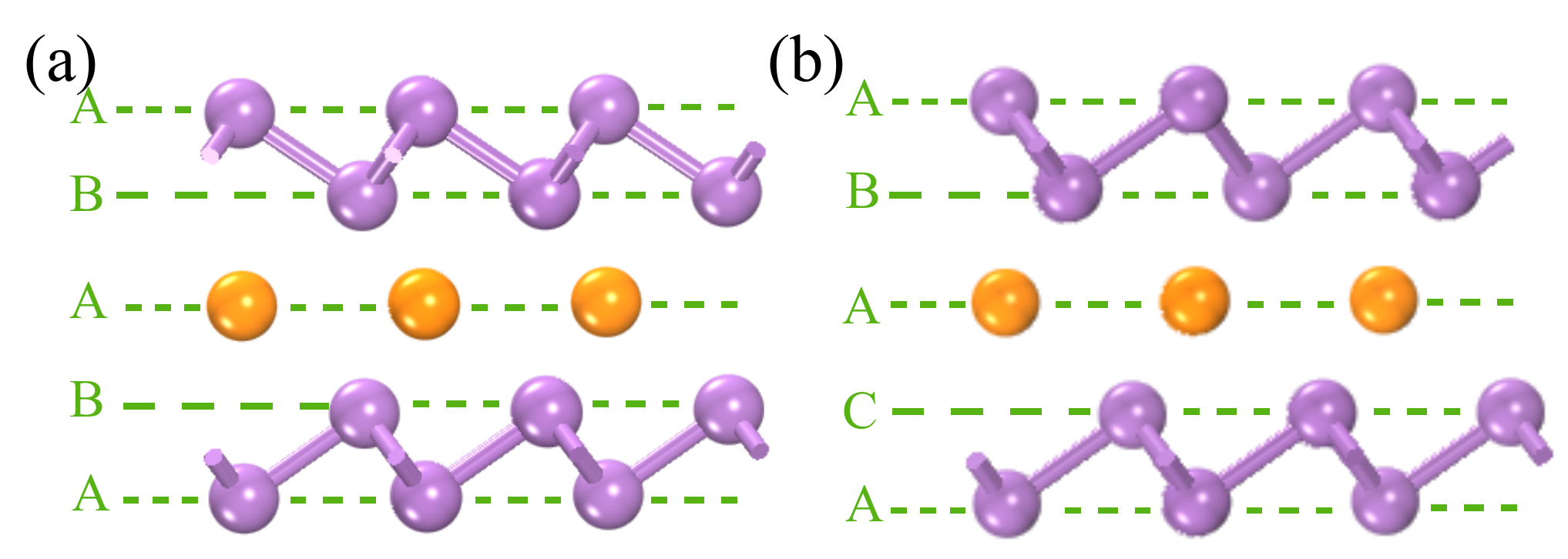}
\caption{Side view of atomic structures of $\beta$-Sb bilayer with Ca-doping. (a) AB-I CG and (b) AB-III CG.}
\label{Fig4}
\end{figure}

As aforementioned, the strong EPC in electron-doped $\beta$-Sb bilayer restricts its doping concentration ($<7.0\times10^{14}$ cm$^{-1}$). In the following, the superconductivity of $\beta$-Sb bilayer under the real Ca-doping conditions will be investigated. Considering the symmetry, there are four mostly possible configurations for Ca intercalation: AB-I CG [Fig.~\ref{Fig4}(a)], AB-II CG [Fig.~S8(a)], AB-III CG [[Fig.~\ref{Fig4}(b)])] and AA CG [Fig.~S8(b)]. After full optimization of all configurations (details are given in Supporting Materials \cite{Supp}), the AB-I CG owns the lowest energy, while the AB-III CG is a little higher in energy ($\sim3.4$ meV/cell) and others are even higher.
%In above subsection, it has been predicted that electron-doped antimonene bilayer maybe a nanoscale superconductor. However, above electron-doping is technically artificial, namely by introducing more electrons into the system with uniform positive charge background for electrical neutrality. In this subsection, the superconductivity of $\beta$-Sb bilayer under the real Ca-doping conditions will be investigated.%

\begin{figure}
\centering
\includegraphics[width=0.4\textwidth]{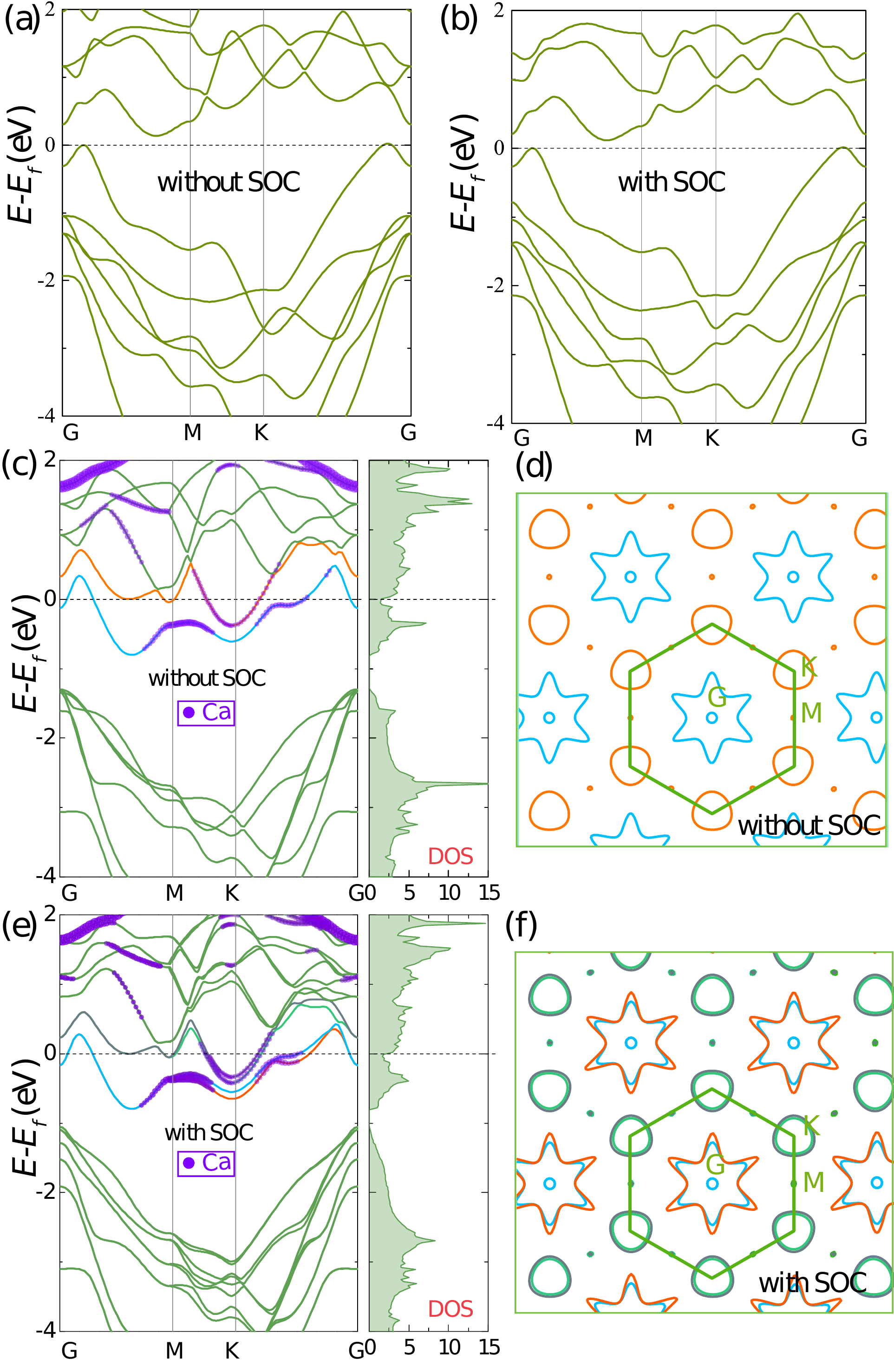}
\caption{Electronic structure of $\beta$-Sb bilayer without (a-b) and with Ca-intercalation (c-f). (c) and (e) Electronic band structures and DOS's of AB-I CG without/with SOC. The violet circles indicate those states with more than $50\%$ of density contributed by Ca. (d) and (f) The corresponding Fermi surfaces.}
\label{Fig5}
\end{figure}

The calculated band structures without/with SOC for AB-I CG (AB-III CG) are shown in Fig.~\ref{Fig5}(c) and (e) (Fig.~S9). As expected, the $\beta$-Sb bilayer transforms from a semimetal to a metal by Ca intercalation. Differing from above pure electron-doped case, there are two bands (blue and orange curves in Fig.~\ref{Fig5}(a)) crossing the Fermi level, and their corresponding Fermi surfaces are shown in Fig.~\ref{Fig5}(b). According to the projection, the hole pocket Fermi surfaces around the $K$ points are mainly contributed by the Ca layer, while the sun-shape Fermi surface around the $\Gamma$ point is constructed by the combination of $\beta$-Sb bilayer and Ca layer. When SOC is considered in calculations for the AB-I CG, the spin degeneracy of the valence and conduction bands along the line $\Gamma-K$ is broken, as shown in Fig.~\ref{Fig5}(e-f). Generally, the Kramer theorem leads to the degeneracy at the $\Gamma$ and $M$ points. For other $k$ points, the spin degeneracy is determined by the time-reversal and $D_{3h}$ point-group symmetry. Due to lack of inversion-reversal symmetry in $D_{3h}$, the spin splitting ($\Delta(k,\varphi)=\beta(k)\left|\sin3\varphi\right|$) is induced at the $K$ point \cite{henk2003spin}, as found in the MoS$_{2}$ monolayer \cite{zhu2011giant}. However, for the AB-III CG (Fig.~S9) with both the inversion-reversal and time-reversal symmetry, SOC can not induce spin splitting in the whole BZ, which is similar to aforementioned electron-doped antimonene bilayer.

\begin{figure}
\centering
\includegraphics[width=0.48\textwidth]{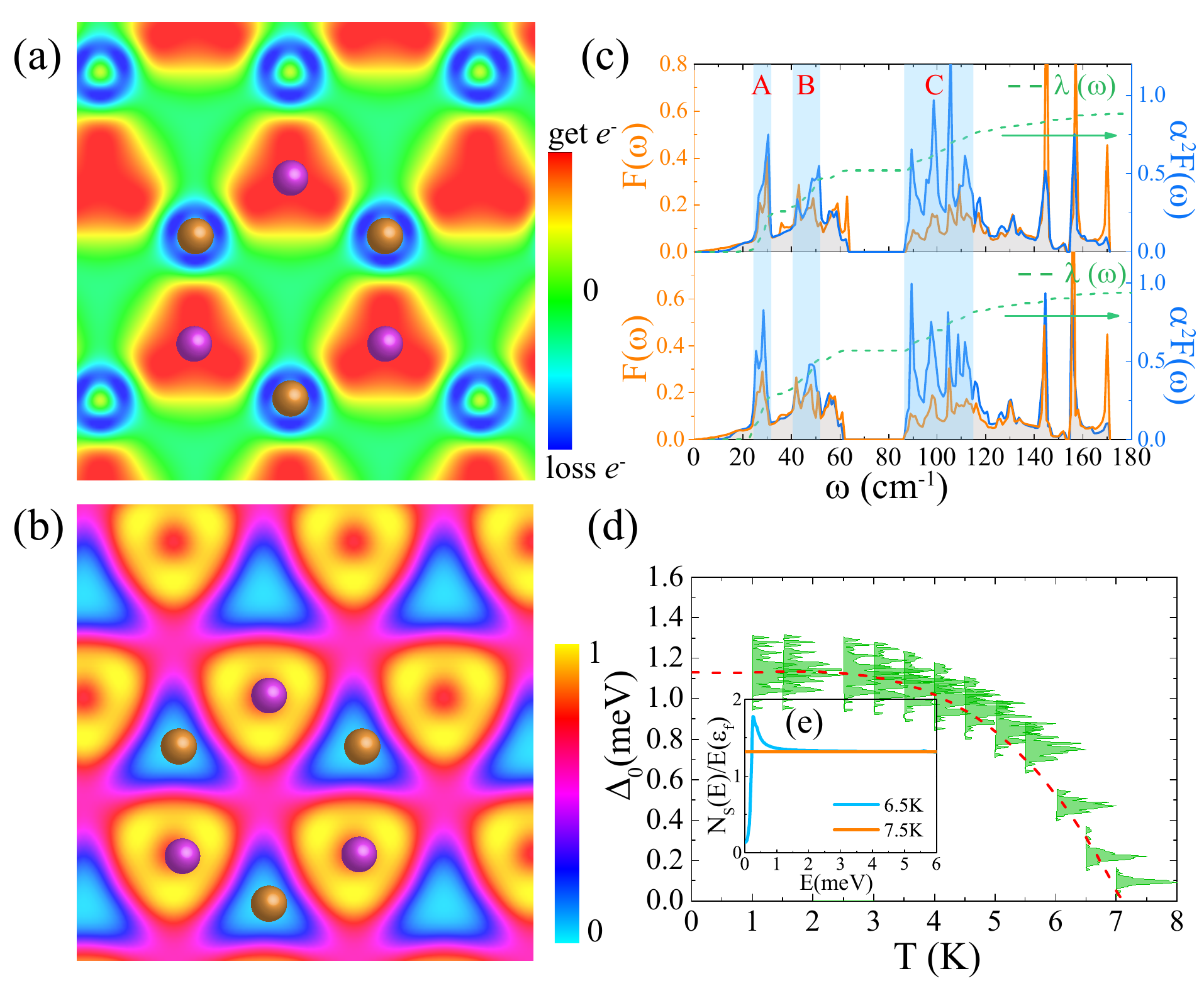}
\caption{Results for $\beta$-Sb bilayer with Ca-intercalation. (a) Electron density difference (between the pristine $\beta$-Sb bilayer and the Ca-intercalated one) viewed along the (001) direction for AB-I CG. (b) Electron localization function along the [001] direction for AB-I CG. Violet: Sb; Yellow: Ca. (c) Phonon density of states (PDOS, $F(\omega)$), Eliashberg-spectral function $\alpha^{2}F(\omega)$, and electron-phonon coupling $\lambda(\omega)$ for AB-I CG. (d) Superconducting gap for AB-I CG. The red curve is the fitting result. (e) Normalized quasiparticle density of states for AB-I in the superconducting and normal state.}
\label{Fig6}
\end{figure}

Like aforementioned electron-doped $\beta$-Sb bilayer, all Raman modes are soften after Ca intercalation (Table~S2) in comparison to pristine $\beta$-Sb bilayer. The charge transfer from Ca layer to Sb layer [Fig.~\ref{Fig6}(a)] results in ionic interaction between Ca and Sb. In addition, the breathing mode $A_{\rm 1g}^1$ of $\beta$-Sb bilayer is not sensitive to the Ca intercalation, differing from other 2D materials (e.g. MoS$_2$ \cite{zhang2016strain}) with full interlayer vdW interaction. Interestingly, through electron localization function (ELF) analysis [Fig.~\ref{Fig6}(b)], the generated Coulomb repulsion between charged Sb-Sb pairs weakens its covalent bond, causing remarkable red shift of resting Raman modes.

\begin{figure}
\centering
\includegraphics[width=0.46\textwidth]{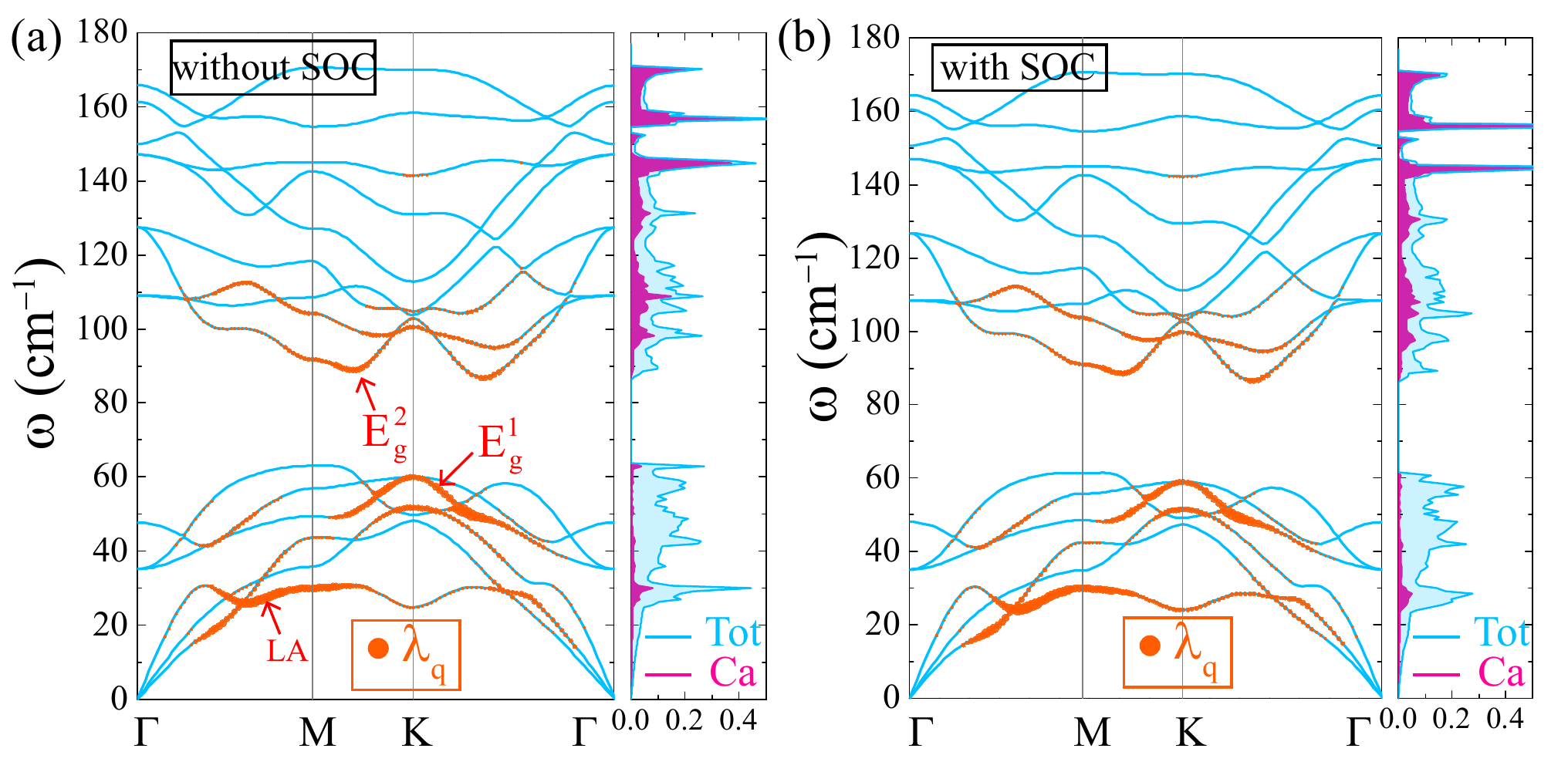}
\caption{Phonon spectrum and electron-phonon coupling of AB-I CG. The orange circles is proportional to the magnitude of electron-phonon couplings $\lambda_{\bf{q}\nu}$.}
\label{FigS9}
\end{figure}

The phonon dispersion and corresponding EPC of AB-I and AB-III are also calculated, as shown in Fig.~\ref{FigS9} and S10. Due to the Fermi contour nesting, the strong EPC is happened around $K$ point, where $E_{\rm g}^1$ and $E_{\rm g}^2$ make great contributions to the EPC. Comparing AB-I with AB-III, the $\lambda_{\bf{q}\nu}$ is obviously different for frequency below $70$ cm$^{-1}$ due to their different stacking and group points. The vibrational modes at low frequency in AB-III CG have stronger EPC than those in AB-I CG.

Our results for $\alpha^{2}F(\omega)$, PDOS, and $\lambda(\omega)$ of AB-I and AB-III CG's are shown in Fig.~\ref{Fig6}(c) and Fig.~S11, respectively. Similar to the pure electron-doping case, $\alpha^{2}F(\omega)$ and $F$($\omega$) of these two have similar peaks, indicating all vibration modes contribute to EPC. Specially, there are three areas (blue shading in Fig.~\ref{Fig6}(c)) where obvious peaks of $\alpha^{2}F(\omega)$ exist in the AB-I CG. These frequency ranges of area-A, area-B, and area-C correspond to the ZA, $E_{\rm g}^1$, and $E_{\rm g}^2$ modes, respectively. Therefore, these three modes promote the superconductivity in Ca-intercalated $\beta$-Sb bilayer. The superconducting $T_{\rm C}$ is also estimated based on Eq.~\ref{McMillan}, which are $4.9$ and $5.3$ K ($6.8$ and $6.9$ K) for the AB-I CG (AB-III CG) without and with SOC, respectively. Thus, the SOC can slightly improve the superconductivity of Ca-intercalated $\beta$-Sb bilayer, particularly for the case without inversion symmetry, which  in agree with the previous study \cite{sklyadneva2013mass}.

From above electronic and phonon calculations, one can notice that the Fermi surface and phonon dispersion are anisotropic due to the remarkable difference between $\Gamma-M$ and $\Gamma-K$, which may raise the question whether the EPC is also anisotropic. Hence, the proper treatment of anisotropic EPC based on Wannier function is adopted in following study. Moreover, the superconducting gap is also calculated, which is given by \cite{margine2013anisotropic}:
\begin{equation}
\Delta(\textbf{k},i\omega_{n})=\frac{\phi(\textbf{k},i\omega_{n})}{Z(\textbf{k},i\omega_{n})},
\label{gap}
\end{equation}
where $Z(\textbf{k},i\omega_{n})$ and $\phi(\textbf{k},i\omega_{n})$ are the mass renormalization function and the order parameter, respectively. From this quantity, the leading edge of the superconducting gap for electron momentum $\textbf{k}$ on the Fermi surface is obtained by solving for $\omega$ in $Re\Delta(\textbf{k},i\omega_{n})=\omega$. The $T_{\rm C}$ of superconductor is obtained from $\Delta_{0}\neq0$ at the highest temperature, while $\Delta_{0}=0$ in the whole temperature range for usual metal. To calculate $\Delta$, the DFT band structure is fitted by using Wannier functions. The obtained superconducting gap of AB-I is shown in Fig.~\ref{Fig6}(d) as a function of temperature. When taking a typical value $\mu^{*}=0.1$, there is nearly only one superconducting gap, which is closed around $7.2$ K. These results indicate the superconducting gap function of Ca-intercalated $\beta$-Sb bilayer is very weakly anisotropic and the corresponding $T_{\rm C}$ is about $7.2$ K, only slightly higher than that estimated using McMillan's approach. To accurately verify the $T_{\rm C}$, real part of the normalized quasiparticle density of states is calculated below and above $T_{\rm C}$, which is based on isotropic approximation:
\begin{equation}
\frac{N_{s}(\omega)}{N(\varepsilon_{F})}=Re[\frac{\omega}{\sqrt{\omega^2-\Delta^2(\omega)}}].
\end{equation}
From Fig.~\ref{Fig6}(e), one can see that Van Hove singularity exists in the normalized quasiparticle density of states at $6.5$ K, which disappears at $7.5$ K. Hence, $T_{\rm C}$ should be between $6.5$ K and $7.5$ K. The ratio between the gap and $T_{\rm C}$ gives: $2\Delta_0(k_{\rm B}T_{\rm C})^{-1}=3.42-3.94$ ($k_{\rm B}$: Boltzmann constant), very close to the ideal BCS value of $3.53$, which is also an evidence for phonon-medium BCS superconductivity in Ca-intercalated $\beta$-Sb bilayer.

In the end of this subsection, the way how to insert Ca into the interlayer of $\beta$-Sb bilayer is proposed. First, it is difficult for Ca to directly cross through the atomic gaps of $\beta$-Sb due to the large energy barrier (about $2.5$ eV according to the NEB result). Even though, there remain two possible ways to insert Ca, i.e. using boundary and vacancy. By inserting Ca from boundary, our MD simulation indicates that Ca atoms could spontaneously go into the interlayer of $\beta$-Sb bilayer. However, due to the stronger interlayer interaction than those in graphene and $h$-BN, there is a large energy barrier (about $1.2$ eV) preventing Ca's freely moving in interior at room temperature. By inserting Ca from vacancy of $\beta$-Sb, there are mainly three processes, including generating vacancy, inserting Ca via vacancy, and restoring vacancy. According to our MD simulation, Ca atoms could cross through the vacancy with the help of thermal fluctuation, but they could not spontaneously move in the interlayer, similar to above boundary-inserting case. Hence, additional driving forces (e.g. electric current) are required for internal moving within the interlayer of $\beta$-Sb bilayer, which deserve further studies.

\subsection{Protective layers}
\begin{figure}
\centering
\includegraphics[width=0.42\textwidth]{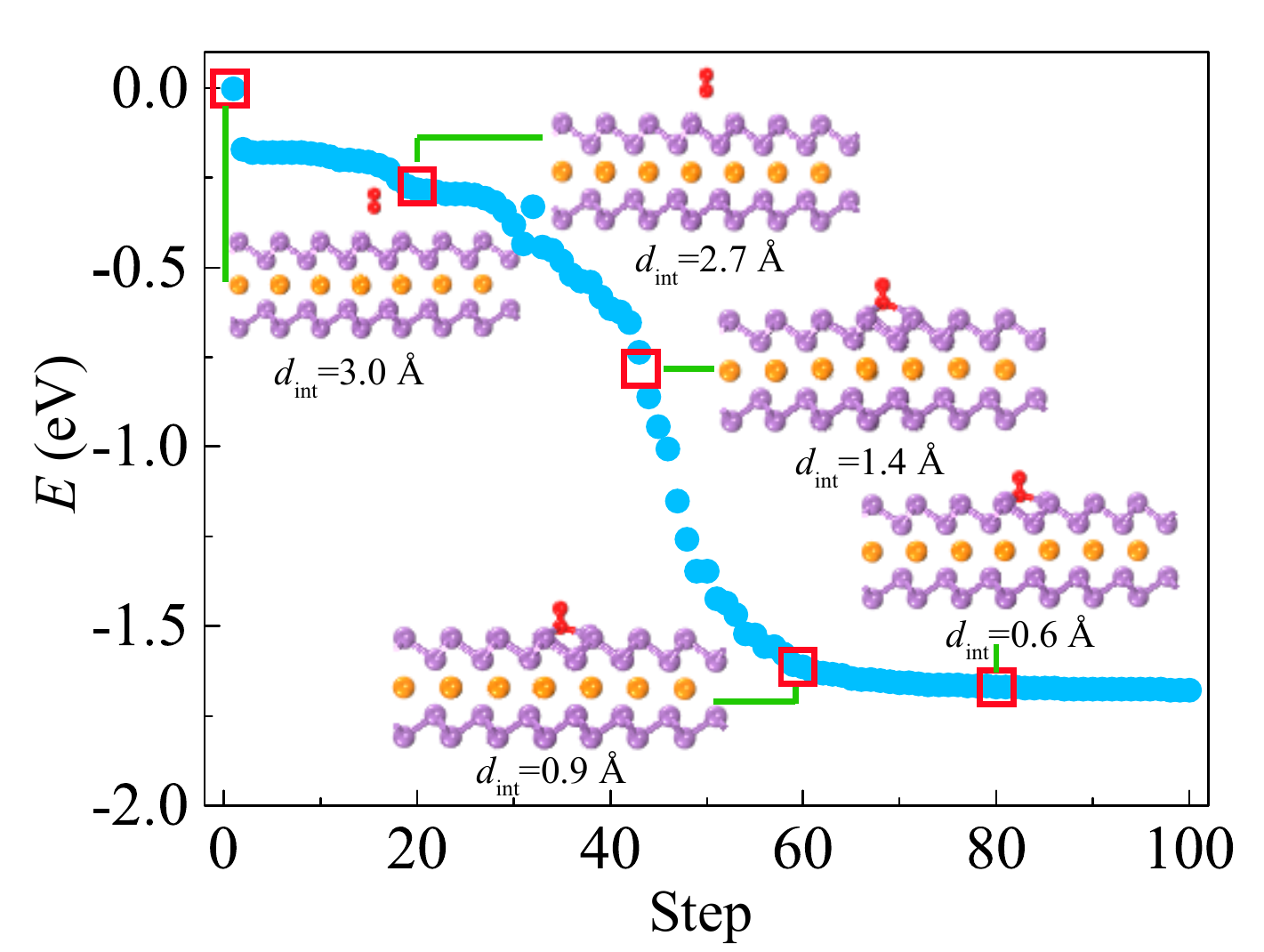}
\caption{MD simulated process of O$_{2}$ adsorption on the AB-I CG surface without energy barrier. The $d_{\rm int}$ indicates the interlayer distance between O$_{2}$ and AB-I CG surface.}
\label{Fig11}
\end{figure}

The superconductivity of $\beta$-Sb bilayer has been predicted upon Ca intercalation. The stability in ambient conditions is the first condition for experiments, which needs careful theoretical investigations. To simulate the Ca-intercalated $\beta$-Sb bilayer in air, here the interaction between O$_{2}$ and AB-I CG is studied using a $4\times4$ supercell. There are three most possibly adsorption sites for O$_{2}$ on its surface, as shown in Fig.~S12, where the C-site owns the lowest energy. The corresponding binding energy ($E_{b}=E_{\rm total}-E_{\rm AB-I}-E_{\rm O_2}$) is calculated to be $-0.9$ eV, indicating strong absorption. The absorption process can be intuitively simulated as the atomic relaxation process in DFT calculation, as shown in Fig.~\ref{Fig11}, suggesting no energy barrier from the physical absorption to chemical absorption.

\begin{figure}
\centering
\includegraphics[width=0.42\textwidth]{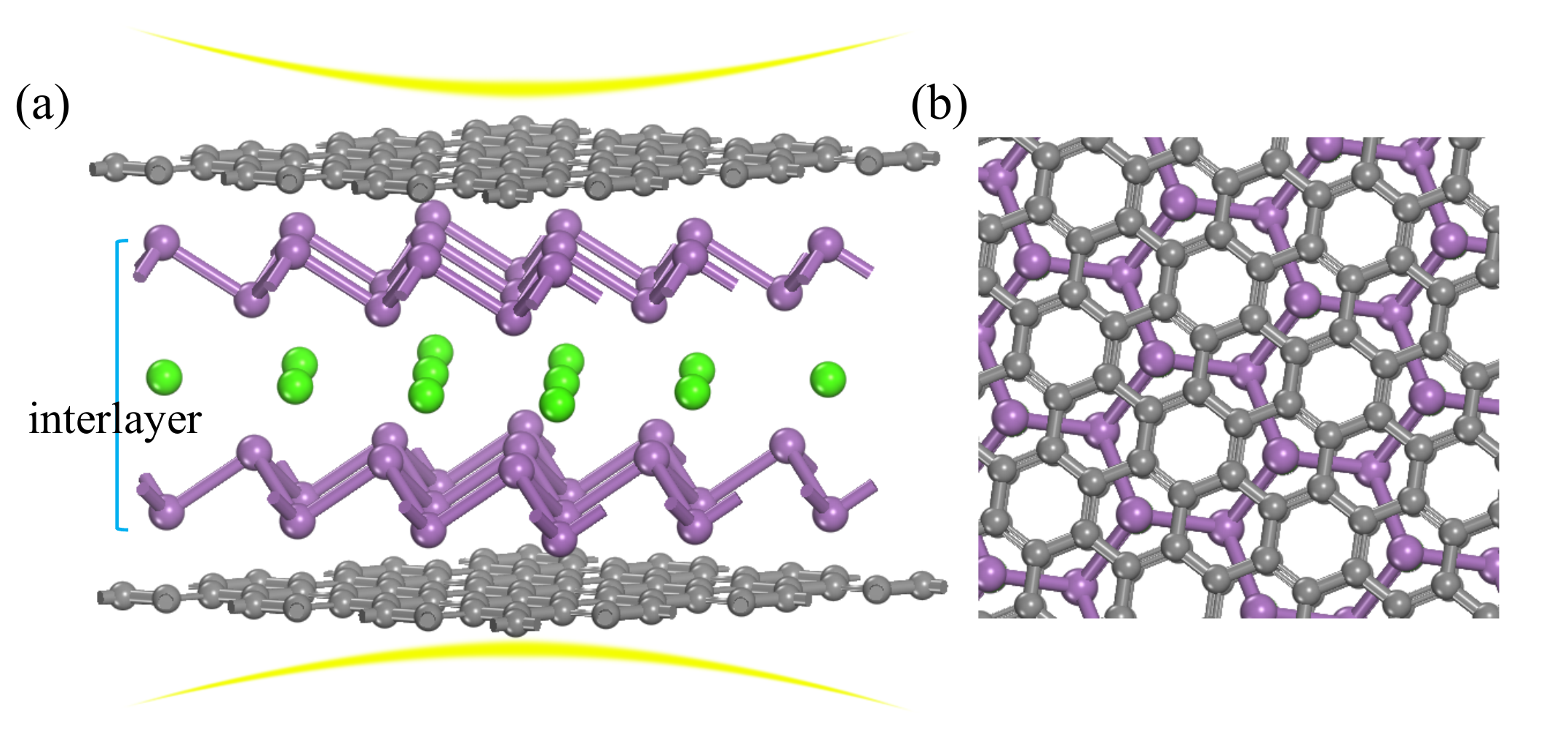}
\caption{(a) Side and (b) top views of atomic structure of AB-I CG with the protective layers.}
\label{Fig8}
\end{figure}

\begin{figure}
\centering
\includegraphics[width=0.48\textwidth]{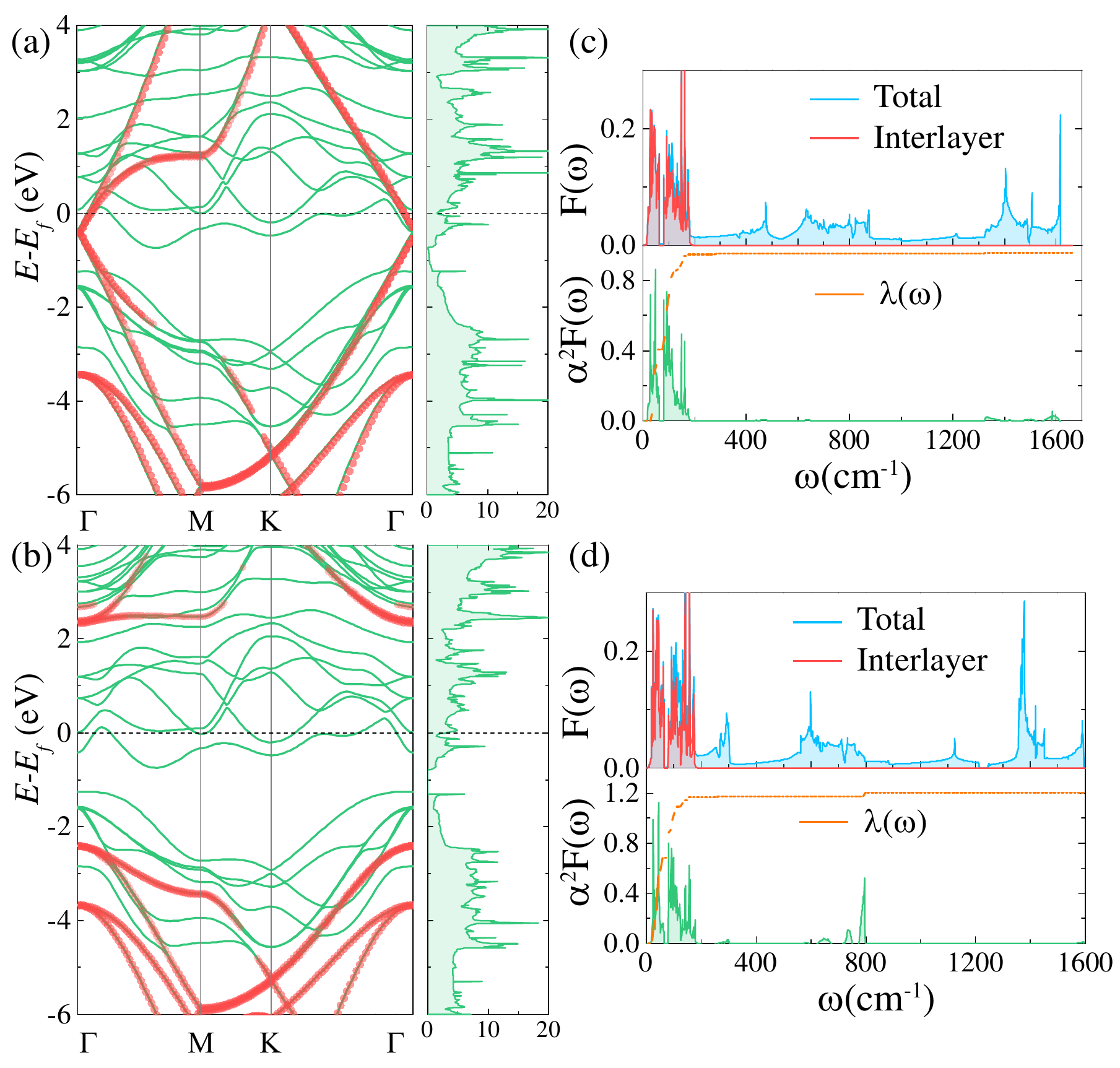}
\caption{Electronic band structures, phonon density of states (PDOS,$F(\omega)$), Eliashberg-spectral function $\alpha^{2}F(\omega)$, and electron-phonon coupling $\lambda(\omega)$ of AB-I CG with two kinds of covered layers. The red circles indicate those states with more than $60\%$ of density contributed by covered layers. (a)(c) The graphene covered bilayer; (b)(d) The $h$-BN covered bilayer.}
\label{Fig9}
\end{figure}

In order to improve the stability and protect the superconductivity of Ca-intercalated $\beta$-Sb bilayer, two kinds of protective layers (graphene and $h$-BN) are considered to avoid the effect of oxygen or water (Fig.~\ref{Fig8}(a)). Considering the physical similarity between AB-I and AB-III CG, here only the AB-I CG is selected as a represent, while the conclusions can be naturally extend to cover AB-III CG. In addition, to reduce the lattice mismatch between interlayer and protective layers, $(\sqrt{3}\times\sqrt{3})R30^{\circ}$ supercell of graphene and $h$-BN are used, as shown in Fig.~\ref{Fig8}(b), where lattice mismatch is less than $2\%$. After full optimization of atomic positions, the distances between the $\beta$-Sb bilayer and protective layers are around $3$ {\AA}, suggesting the vdW-type coupling. The corresponding electronic structure for graphene- and $h$-BN covered $\beta$-Sb layers are shown in Fig.~\ref{Fig9}(a) and ~\ref{Fig9}(b), respectively. Due to the insulating nature of $h$-BN, it has negligible effect on the Fermi surface of Ca-intercalated $\beta$-Sb bilayer. However, the graphene's Dirac cone will affect the Fermi surface around the $\Gamma$ point.

The dynamic stability of these two systems are also confirmed by phonon dispersion [Fig.~S13 (a-b)], and the occupied phonon states of protective layers are mainly in high frequency region due to their small masses and strong $sp^2$ covalent bonds. To verify the effect of protective layers on the superconductivity, the corresponding $\alpha^{2}F(\omega)$, PDOS, and $\lambda(\omega)$ are also calculated, as shown in Fig.~\ref{Fig9}(c-d). As expected, the superconductivity is mainly concentrated at the interlayer, while the impact of covered layers are tiny according to the shape of $\alpha^{2}F(\omega)$ and PDOS. The obtained $T_{\rm C}$ based on McMillan's approach are $7.3$ and $9.6$ K for the graphene- and $h$-BN covered layers, respectively. The improvement of $T_{\rm C}$ may be part from lattice mismatch (about $1.6\%$ tensile stress). To verify this guess, the strain effect on pure Ca-intercalated $\beta$-Sb bilayer is also calculated, which gives $T_{\rm C}$ up to $6.7$ K. In this sense, the covered layers may also partially participate in EPC, besides the pure strain effect. In particular, the superconductive properties of $h$-BN covered bilayer is superior than the graphene covered one, due to less disturbance on Fermi surface, making it a promising candidate as a nanoscale superconductor.

\section{Conclusion}
In summary, using the first-principles calculations, it is found that different stacking leads to varying interlayer interaction and interlayer hopping parameter, which have great influence on electronic structure of $\beta$-Sb bilayer. In addition, electron-doping and Ca-intercalation can transform the $\beta$-Sb bilayer from a semimetal to a superconductor. However, the strong EPC would lead to dynamic instability of $\beta$-Sb bilayer, when the doping density is $>0.8$ electrons/cell. Moreover, the stability of antimonene bilayer in air is decreased after the electron transfer. For further experiments and applications, two kinds of protective layers (graphene and $h$-BN) have been designed, which have little influence on the Fermi surface of Ca-intercalated $\beta$-Sb bilayer. More importantly, the superconducting $T_{\rm C}$ is enhanced to $9.6$ K with the help of these protective layers.

\acknowledgments{Work was supported by the National Natural Science Foundation of China (Grant Nos. 11674055, 11834002), the Scientific Research Foundation of Graduate School of Southeast University (Grant No. YBJJ1769), and Fundamental Research Funds for the Central Universities. Most calculations were done on Tianhe II of National Supercomputer Center in Guangzhou (NSCC-GZ)}

\bibliographystyle{apsrev4-1}
\bibliography{ref}

\end{document}